\documentclass[showpacs,twocolumn,prl]{revtex4-1}
\usepackage[utf8]{inputenc}
\usepackage{graphicx}
\usepackage{color}
\usepackage{appendix}
\usepackage{amsmath}
\usepackage{amssymb}
\usepackage{subfigure}
\usepackage{epsfig}

\newcommand{\om}{\Omega}
\newcommand{\omdag}{{\Omega}^{\dagger}}
\newcommand{\omeq}{\Omega_{\rm eq}}

\newcommand{\delom}{\delta\Omega}

\newcommand{\F}{\boldsymbol{F}}

\newcommand{\p}{\boldsymbol{p}}
\newcommand{\vv}{\boldsymbol{v}}
\newcommand{\rr}{\boldsymbol{r}}
\newcommand{\nab}{\boldsymbol{\nabla}}
\newcommand{\xxi}{\boldsymbol{\xi}}

\newcommand{\eeta}{\boldsymbol{\eta}}
\newcommand{\R}{\boldsymbol{R}}

\begin{document}

\title{Green-Kubo approach to the average swim speed in active Brownian systems}

\author{A.~Sharma}
\author{J.M.~Brader}
%\email{joseph.brader@unifr.ch}
\affiliation{Department of Physics, University of Fribourg, CH-1700 Fribourg, Switzerland}


\begin{abstract}
We develop an exact Green-Kubo formula relating nonequilibrium averages in systems of 
interacting active Brownian particles to equilibrium time-correlation functions. 
%Linear response relations emerge as a special case for low activity. 
The method is applied to calculate the density-dependent average swim speed, which is a 
key quantity entering coarse grained theories of active matter. 
The average swim speed is determined by integrating the equilibrium autocorrelation 
function of the interaction force acting on a tagged particle. 
Analytical results are validated using Brownian dynamics simulations. 
%
%
%
%The autocorrelation is evaluated for various interaction potentials using equilibrium 
%Brownian dynamics simulations.  
%
\end{abstract}

%\pacs{\joe64.70.pv, 64.70.Q-, 83.80.Ab, 83.60.La}

\maketitle

Assemblies of active, interacting Brownian particles (ABPs) are intrinsically nonequilibrium systems. 
In contrast to equilibrium, for which the statistical mechanics of Boltzmann and Gibbs 
enables the calculation of average properties, there is no analogous framework 
out-of-equilibrium. 
However, useful exact expressions exist, which enable average quantities to be calculated 
in the nonequilibrium system 
by integrating an appropriate time correlation function; the Green-Kubo formulae of linear response theory 
\cite{green,kubo,kubo_book}. Transport 
coefficients, such as the diffusion coefficient or shear viscosity, are thus conveniently related 
to {\it equilibrium} autocorrelation functions.    
%
%Familiar examples are the diffusion coefficient as an integral over the velocity autocorrelation 
%function or the viscosity as an integral over shear stress fluctuations. 
%
Given the utility of the approach it is surprising that the application of Green-Kubo-type 
methods to active Brownian systems has received little attention \cite{seifert}. 

The primary aim of the present work is to extend Green-Kubo-type methods to treat ABPs. 
This approach has two appealing features. Firstly, information about the active system 
can be obtained from equilibrium simulations. 
%thus enabling 
%identification of the relevant time-scales. 
Secondly, the exact expressions derived provide a solid starting point for the development 
of approximation schemes and first-principles theory. 
The method we employ is a variation of the integration-through-transients 
approach, originally developed for treating interacting Brownian particles subject to external 
flow \cite{evans_morriss,fuchscates2002,fuchscates2005,brader2012}. 

A fundamental feature of ABPs is the persistent character of the particle trajectories. 
For strongly interacting many-particle systems the interplay between persistent motion and 
interparticle interactions can generate a rich variety of collective 
phenomena, such as motility-induced phase separation (see \cite{cates_tailleur_review} for 
a recent overview). A quantity which features prominantly in many theories of ABPs 
\cite{cates_tailleur_review,cates_pnas,fily_marchetti,speck_lowen,krinninger_schmidt_brader} is 
the density-dependent average swim speed, which describes how the motion of each particle is obstructed 
by its neighbours. 
Given the ubiquity of the average swim speed in the literature on ABPs we choose it 
as a relevant observable with which to illustrate our general Green-Kubo-type approach. 
We  demonstrate that this quantity can be obtained from a history integral over the 
equilibrium autocorrelation of 
tagged-particle force fluctuations, which we investigate in detail using Brownian dynamics 
(BD) simulation. 

We consider a three dimensional system of $N$ active, interacting, spherical Brownian particles 
with coordinate $\rr_i$ and orientation specified by an embedded unit vector $\p_i$. 
A time-dependent self-propulsion of speed $v_0(t)$ 
acts in the direction of orientation. 
Allowing for time-dependence of this quantity both clarifies the general structure of the 
theory and leaves open the possibility to model physical systems for which the the amount 
of fuel available to the particles is not constant (see e.g.~\cite{vt1,vt2}). 
Omitting hydrodynamic interactions the motion can be modelled by the Langevin equations
\begin{align}\label{full_langevin}
&\!\!\!\!\!\!\dot{\rr}_i = v_0(t)\,\p_i  + \gamma^{-1}\F_i + \xxi_i\;\;,
\;\;\;
\dot{\p}_i = \eeta_i\times\p_i \,,
\end{align}
where $\gamma$ is the friction coefficient and the force on particle $i$ is generated from the 
total potential energy according to $\F_i\!=\!-\nabla_i U_N$.
The stochastic vectors $\xxi_i(t)$ and $\eeta_i(t)$ are Gaussian distributed with zero mean and 
have time correlations	
$\langle\xxi_i(t)\xxi_j(t')\rangle=2D_t\boldsymbol{1}\delta_{ij}\delta(t-t')$ and 
$\langle\eeta_i(t)\eeta_j(t')\rangle=2D_r\boldsymbol{1}\delta_{ij}\delta(t-t')$. 
The translational and rotational diffusion coefficients, $D_t$ and $D_r$, are treated 
in this work as independent model parameters.

It follows exactly from \eqref{full_langevin} that the  
joint probability distribution, $\!P(\rr^N\!\!,\p^N\!\!,t)$, evolves according to 
\cite{gardiner}
\begin{align}\label{smol_eq}
\frac{\partial P(t)}{\partial t} = \om_{\rm a}(t) P(t).
\end{align}
The time-evolution operator can be split into a sum of two terms, $\om_{\rm a}(t)=\omeq+\delom_{\rm a}(t)$, 
where the equilibrium contribution is given by 
\begin{align}\label{smol_op_eq}
\omeq = \sum_{i=1}^{N} \nab_{i}\!\cdot\!
\big[
D_\text{t}\!\left(\nab_{i}\! - \beta\F_i\right) 
\big] \!+\! D_\text{r}\R_i^2, 
\end{align}
with rotation operator $\R\!=\!\p\times\!\nabla_{\!\p}$ (see, e.g.~\cite{morse_feshbach}) 
and the time-dependent, active part of the dynamics is described by the operator  
\begin{align}\label{smol_op_del}
\delom_{\rm a}(t) = 
-\sum_{i=1}^{N} v_0(t)\nab_{i}\!\cdot\p_i. 
\end{align}

%In \eqref{smol_op_eq} and \eqref{smol_op_del} all differential operators act on everything 
%to the right.
%We now employ \eqref{smol_eq} as a basis for deriving generalized Green Kubo 
%formulae. 
%Firstly, 
To solve \eqref{smol_eq} we define a nonequilibrium part of the distribution 
function, $\delta P(t) = P(t) - P_{\rm eq}$ \cite{fuchs_cates}, 
%
%\begin{align}\label{Psplit}   
%\delta P(t) = P(t) - P_{\rm eq}\,, 
%\end{align}
%
where $P_{\rm eq}$ is the equilibrium distribution of position and orientation. 
Using $\omeq P_{\rm eq}=0$ yields the equation of motion 
\begin{align}\label{delP_eom}   
\frac{\partial}{\partial t}\delta P(t) = \om_{\rm a}(t) \delta P(t) 
+ \delom_{\rm a}(t) P_{\rm eq}\,. 
\end{align}
Treating the last term as an inhomogeneity and solving for $\delta P(t)$ 
we obtain a formal solution for the nonequilibrium distribution
\begin{align}\label{P_eom}   
P(t) = P_{\rm eq} \,-\, \int_{-\infty}^{t}\!dt' 
v_0(t')
\,e_{+}^{\int_{t'}^{t}ds\,\om_{\rm a}(s)}
\beta F^p P_{\rm eq}\,,
\end{align}
where $e_{+}(\cdot)$ is a positively ordered exponential function 
(see the appendix in \cite{brader2012}) and we have used 
$\delom_{\rm a}(t)P_{\rm eq} = -\beta v_0(t)F^p P_{\rm eq}$, with `projected force'
fluctuation 
\begin{align}
F^p = \sum_i \p_i\cdot\F_i.
\end{align}
%
%\begin{align}\label{Fp}
%F^p = \sum_i \p_i\cdot\F_i. 
%\end{align}
%
The projected force emerges as a central quantity within our approach and 
indicates to what extent the interparticle interaction forces act in the direction of 
orientation, either assisting or hindering the self-propulsion. 
We will show that this quantity is closely related to the average swim speed 
in the active system. 

Introducing a test function, $f$, on the space of positions and orientations 
and integrating \eqref{P_eom} by parts yields a formally exact 
expression for a nonequilibrium average 
\begin{align}\label{nonlinear_average}
\langle f \rangle(t) = \langle f \rangle_{\rm eq} - \int_{-\infty}^{t}\!dt'\,
v_0(t')\langle \beta F^p e_{-}^{\int_{t'}^{t}ds\,\omdag_{\rm a}(s)}f \rangle_{\rm eq}, 
\end{align}
where $e_{-}(\cdot)$ denotes a negatively ordered exponential 
\cite{brader2012} and $\langle\cdot\rangle_{\rm eq}$ is an equilibrium average 
over positional and orientational degrees of freedom. 
The adjoint operator is given by 
$\omdag_{\rm a}(t)=\omdag_{\rm eq}-\delom_{\rm a}(t)$, where 
\begin{align}
\omdag_{\rm eq} = \sum_{i} 
D_\text{t}\!\left(\nab_{i}\! + \beta\F_i\right) 
\!\cdot\!\nab_{i} \!+\! D_\text{r}\R_i^2
\end{align}
generates the equilibrium dynamics.
%
%\begin{align}\label{adj_smol}
%\omdag(t) = \sum_{i=1}^{N} 
%D_\text{t}\!\left(\nab_{i}\! + \beta\F_i\right) 
%\!\cdot\!\nab_{i} \!+\! D_\text{r}\R_i^2 
%+\nab_{i}\!\cdot\!
%\left(v_0(t)\,\p_i\right).  
%\end{align}
%
%%Equation \eqref{nonlinear_average} is a central result of the present work. 
The integrand appearing in \eqref{nonlinear_average} involves the {\it equilibrium} 
correlation between the projected force at time $t'$ and 
the observable $f$, which evolves from $t'$ to $t$ according to 
the full dynamics.  
%which makes very convenient the construction of linear response formulae: 
The average is nonlinear in $v_0(t)$, because of the activity 
dependence of the adjoint operator. 

%{\bf Linear response.} 
The response of the system to linear order in $v_0(t)$ is obtained by replacing 
the full time-evolution operator $\omdag_{\rm a}(t)$ in \eqref{nonlinear_average} 
by the time-independent equilibrium operator $\omdag_{\rm eq}$. 
%We thus obtain
%%
%\begin{align}\label{linear_tdep}
%\!\!\langle f \rangle_{\rm lin}(t) = \langle f \rangle_{\rm eq} -\! \int_{-\infty}^{t}\!dt'\,
%v_0(t')\langle \beta F^p e^{\omdag_{\rm eq}(t-t')} f \rangle_{\rm eq}.
%end{align}
%% 
%This expression requires as input an equilibrium time-correlation function. 
Further simplification occurs if the activity is constant in time, $v_0(t)\rightarrow v_0$, 
leading to
\begin{align}\label{linear_tindep}
\langle f \rangle_{\rm lin} = \langle f \rangle_{\rm eq} - v_0 \int_{0}^{\infty}\!dt\,
\langle \beta F^p e^{\omdag_{\rm eq}t} f \rangle_{\rm eq}, 
\end{align}
which can be used to define a general active transport coefficient 
$\alpha\!=\!\lim_{v_0 \to 0}(\langle f \rangle_{\rm lin}\!-\! \langle f \rangle_{\rm eq})/v_0$. 
Equation \eqref{linear_tindep} is the desired Green-Kubo relation for calculating the linear 
response of ABPs to a time-independent activity. 
%%
%%
%%----Here is the second order correction, in case we need it.
%%
%%Expanding \eqref{nonlinear_average} to second order in activity gives the leading correction 
%%to linear response
%%%
%%\begin{eqnarray}
%%\!\!\!\!\!\!\!\!\langle f \rangle = 
%%\langle f \rangle_{\rm lin} - v_0^2\int_{0}^{\infty}\!dt\,
%%\langle \beta F^p \delta\tilde{\Omega}(t)\, e^{\omdag_{\rm eq}t}  f \rangle_{\rm eq}, 
%%\end{eqnarray}
%%
%%where $\delta\tilde{\Omega}(t) = v_0^{-1}\int_0^{t}\!ds\,e^{\omdag_{\rm eq}s}\delomdag_{\rm a} 
%%e^{-\omdag_{\rm eq}s}$. 

As mentioned previously, a quantity of current interest is the average, 
density-dependent swim speed, $v(\rho)$. This describes how the bare swim 
speed, $v_0$, is influenced by interparticle interactions and is an important quantity in many 
of the various theories addressing ABPs 
\cite{cates_tailleur_review,cates_pnas,fily_marchetti,speck_lowen,krinninger_schmidt_brader}.
In particular, the tendency of the system to 
undergo motility-induced phase-separation is determined by the rate of decrease of $v(\rho)$ 
with increasing density; a positive feedback mechanism can result when increasing the local 
density leads to a sufficiently strong reduction of the local
average swim velocity.
%This is found (via simulation) to be a 
%decreasing function of density, starting from its low density limit, $v(\rho)=v_0$, where the 
%particles can move unhindered. 
%
%
\begin{figure}
\begin{minipage}{0.81\textwidth}
\hspace*{-10.4cm}\includegraphics[width=0.31\linewidth,height=0.17\textheight]{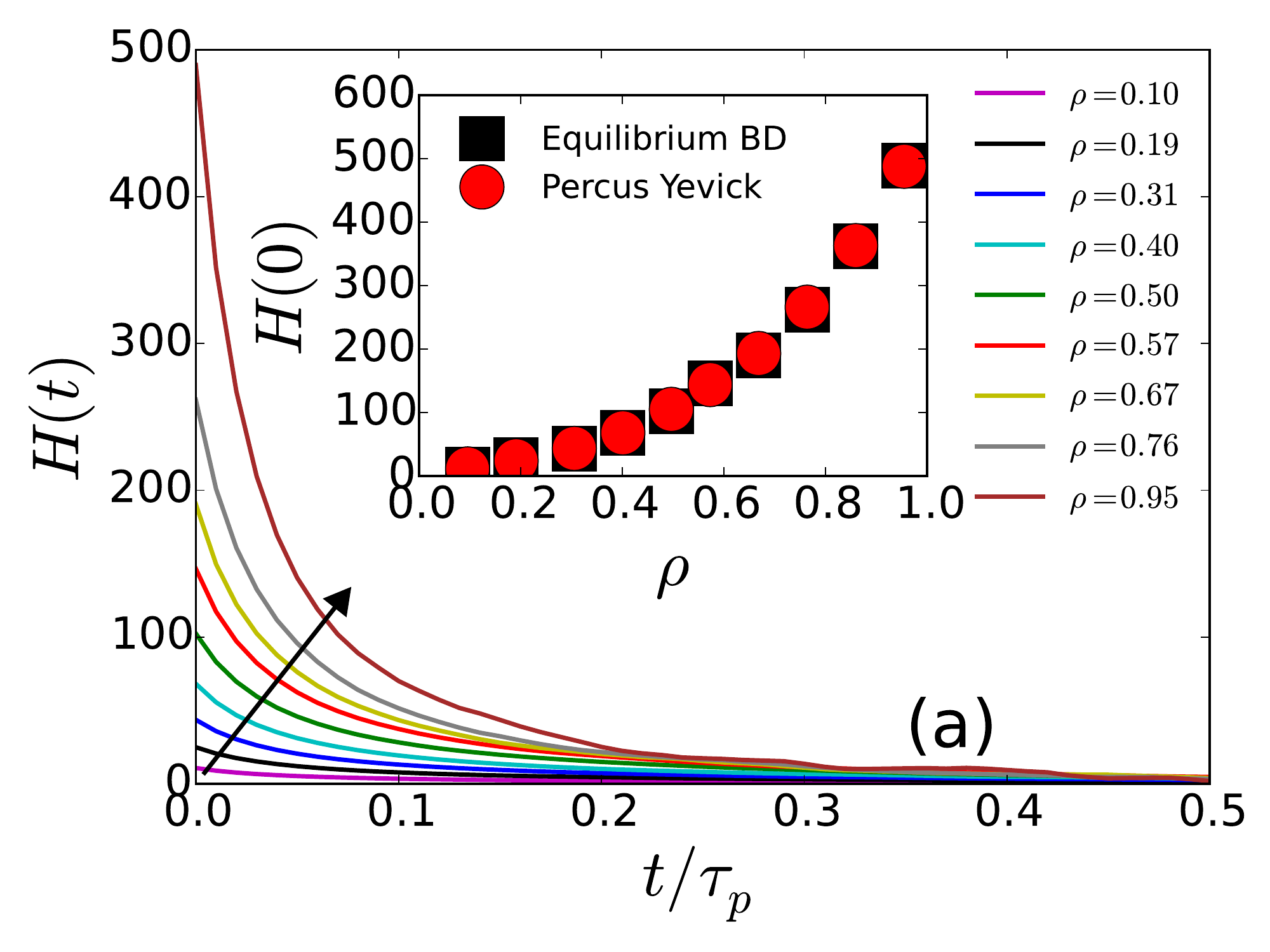}
\end{minipage}
\begin{minipage}{0.81\textwidth}
\vspace*{-4.0cm}\hspace*{-1.7cm}\includegraphics[width=0.31\linewidth,height=0.17\textheight]{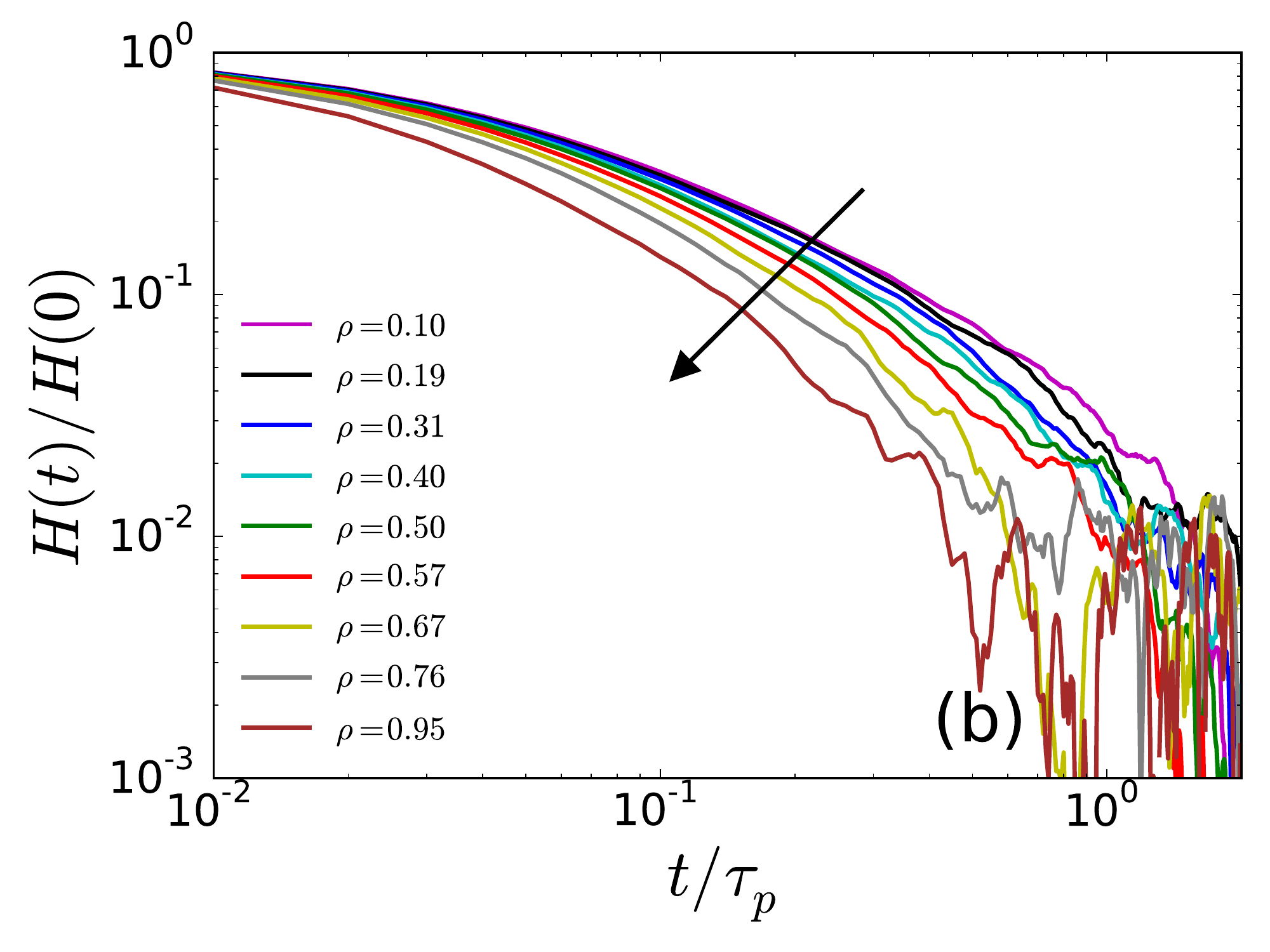}
\end{minipage}
\vspace*{0.0cm}
\caption{\label{fig:correlator}
(a) The correlator $H(t)$ for a system of soft spheres. 
%(truncated Lennard-Jones, $\varepsilon=1.0$). 
The arrow indicates the direction of 
increasing density. Inset: The initial value $H(0)$ as a function of the density. 
Squares: BD simulation. Circles: using equation \eqref{yvon_result} and the 
Percus-Yevick $g_{\rm eq}(r)$ as input.
%volume fraction, $\phi\!=\!\pi\sigma^3\!\rho/6$. 
(b) The same data as in (a) on a log scale. The relaxation 
becomes faster with increasing density. 
}
\end{figure}
The average swim speed is defined as the nonequilibrium average
\begin{align}\label{swim_speed_definition}
v(\rho)=\frac{1}{N}\left\langle \sum_i \vv_i \cdot \p_i \right\rangle
\end{align}
where $\vv_i$ is the velocity of particle $i$. 
Using \eqref{full_langevin} to eliminate the velocity in favour of the forces and using the 
fact that the Brownian force $\xxi_i$ is uncorrelated with the orientation $\p_i$, 
it follows that 
\begin{align}\label{swim_speed_Fp}
v(\rho) = v_0 + \frac{\gamma^{-1}}{N}\langle F^p \rangle. 
\end{align}
For a time-independent $v_0$ we can employ \eqref{linear_tindep} to calculate the 
average in \eqref{swim_speed_Fp} to linear order 
\begin{align}\label{GK_vrho}
v(\rho) = v_0\left(
1 - D_t\int_0^{\infty}\!dt\, H(t)
\right)\,,
\end{align}
where the integrand is the equilibrium autocorrelation of projected force fluctuations
\begin{align}\label{correlator}
H(t) = \frac{1}{N}
\langle
\,\beta F^p e^{\omdag_{\rm eq}t} \beta F^p
\,\rangle_{\rm eq}\,.
\end{align}
Spatial and orientational degrees of freedom decouple in equilibrium, which enables 
the orientational integrals in \eqref{correlator} to be evaluated exactly. 
This yields  
\begin{align}\label{correlator_int}
H(t)
= \frac{1}{3}e^{-2D_r t}\beta^2
\left\langle \F\cdot e^{\omdag_{\rm eq,s}t}\F \right\rangle_{\rm eq, s}, 
\end{align}
where $\F$ is the interaction force acting on an arbitrarily chosen (`tagged') particle, 
$\omdag_{\rm eq,s} \!= \sum_{i} D_\text{t}\!\left(\nab_{i}\! + \beta\F_i\right) 
\!\cdot\!\nab_{i}$ is the spatial part of the time-evolution operator and 
$\langle\cdot\rangle_{\rm eq,s}$ indicates an equilibrium average over spatial degrees 
of freedom. 
The initial value is given by $H(0)\!=\!\beta^2\langle |\F|^2 \rangle_{\rm eq}/3$. 
If we consider pairwise additive interaction potentials, then the Yvon theorem 
\cite{HM} leads to
\begin{align}\label{yvon_result}
H(0)=\frac{1}{3}\,\rho\! \int \!d\rr\,g_{\rm eq}(r)\nabla^2\beta u(r)\,,
\end{align}
where $\rho$ is the number density, 
$u(r)$ is the passive pair potential and $g_{\rm eq}(r)$ is the corresponding 
equilibrium radial distribution function.

%$H(t)$ vanishes in the low density limit, leading to the expected result 
%$v(0)\!=\! v_0$. 

Equation \eqref{correlator_int} shows that the nontrivial physics underlying the linear 
response of the system to activity is contained in the tagged-particle force-autocorrelation 
function. 
This function was encountered many years ago by Klein and coworkers 
\cite{klein} in a study of the velocity autocorrelation in overdamped 
Brownian systems. 
By manipulation of the operator \eqref{smol_op_eq} it was shown that
\begin{align}\label{klein_result}
\left\langle \F(t)\cdot \F(0) \right\rangle_{\rm eq, s} \!\!=\!
\frac{3}{(\beta D_t)^2}\big(
D_t\,\delta(t) - Z_{\rm eq}(t)
\big) ,
\end{align}
where $Z_{\rm eq}(t)$ is the velocity autocorrelation function, defined in terms of the tagged 
particle velocity, ${\bf v}(t)$, according to the familiar relation
\begin{align}
Z_{\rm eq}(t)=\frac{1}{3}\left\langle {\bf v}(t)\cdot{\bf v}(0) \right\rangle_{\rm eq,s}. 
\end{align}
The velocity autocorrelation function is a quantity of fundamental interest in describing the 
dynamics of interacting liquids and is closely related to other important quantities 
(e.g.~the mean-squared displacement and self diffusion coefficient).   
Substituting \eqref{klein_result} into \eqref{correlator_int} yields 
\begin{align}\label{H_vct}
H(t) = \frac{1}{D_t^2}e^{-2 D_r t}\big(
D_t\,\delta(t) - Z_{\rm eq}(t)
\big),
\end{align}
thus providing, via \eqref{GK_vrho}, a direct connection between $Z_{\rm eq}(t)$ and 
$v(\rho)$. 
%which will be discussed further below, describes how the interaction force acting on a tagged 
%particle decorrelates as a result of configurational relaxation. 
The latter can thus be determined to linear 
order in $v_0$ using a standard, equilibrium BD simulation. 
Finally, we note that $H(t)$ remains integrable in all spatial dimensions, because of the 
exponential in \eqref{correlator_int}. 
There is thus no principal difficulty in calculating $v(\rho)$ in two dimensions, in contrast 
to the situation for transport coefficents, such as the self-diffusion coefficient, for which 
the relevant Green-Kubo time-integral diverges \cite{evans_morriss}.

In a recent study of the pressure in active systems Solon {\it et al.} \cite{solonPRL} 
express the density-dependent average swim speed in the form $v(\rho)=v_0 + I_2/\rho$, 
where $\rho$ is the bulk number density. The interaction potential is encoded in the quantity 
$I_2$ via its dependence on a {\it static} structural correlation between density and 
polarization, which are given, respectively, by the first and second harmonic moments of the 
orientation-resolved single particle density. 
This leads to the identification $I_2=-D_t\rho\, v_0\int_0^{\infty}dt\, H(t)$.  
An advantage of the present Green-Kubo formulation over that of Solon {\it et al.} 
is that it enables identification of the relevant relaxation processes contributing to 
the decrease of $v(\rho)$. 
Moreover, we anticipate that \eqref{GK_vrho} will prove more convenient for the development 
of approximations. 

\begin{figure}
\includegraphics[width=0.85\columnwidth]{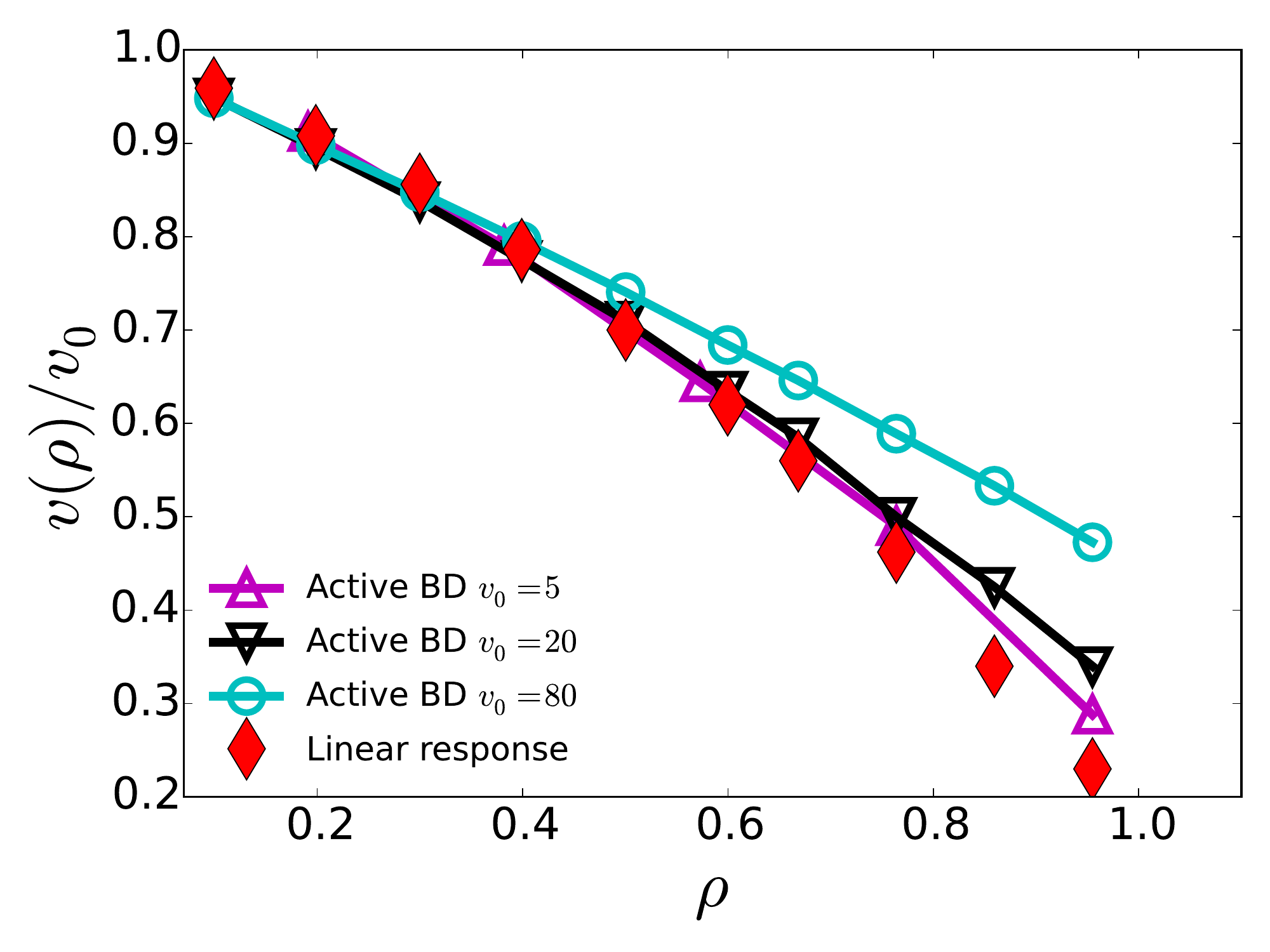}
\vspace*{0cm}
\caption{\label{fig:projected_velocity}
The scaled average swim speed as a function of density.
%volume fraction for $v_0=5, 20$ and $80$. 
Lines with symbols: data from direct calculation of \eqref{swim_speed_definition} using 
active BD simulations. 
Diamonds: the linear response result (independent of $v_0$) calculated 
using the equilibrium time correlation function data from Fig.\ref{fig:correlator} as input 
to \eqref{GK_vrho}.  
}
\label{fig1}
\end{figure}
%

%Comments regarding the simulation of $H(t)$. It can be obtained from equilibrium BD simulations. 
%It does not have any problems with defining the velocity:~$F^p$ is a perfectly well-defined quantity. 

In order to test the range of validity of the linear response result \eqref{GK_vrho} 
we perform BD simulations on a three-dimensional system of $N\!=\!1000$ particles 
interacting via the pair-potential 
$\beta u(r) = 4\varepsilon((\sigma/r)^{12} - (\sigma/r)^6)$, 
where $\sigma$ sets the length scale and we set $\varepsilon=1$. 
The potential is truncated at its minimum, $r=2^{1/6}\sigma$ to yield a softly repulsive 
interaction. 
%In our simulations, 
The system size $L$ is determined as $L\!=\!(N/\rho)^{1/3}$ in order to obtain the desired 
density.
%volume fraction, $\phi\!=\!\pi\sigma^3\!\rho/6$.  
The integration time step is fixed to $dt\!=\!10^{-5} \tau_B$ where $\tau_B\!=\!d^2/D_t$ is 
the time-scale 
of translational diffusion. 
%For any $\rho$, 
Measurements are made after a minimum time of $20\tau_B$ to 
ensure equilibration. In order to measure time-correlations the system is sampled 
every $\tau_p/100$ s, where $\tau_p = 1/2D_R$ is the rotational diffusion time scale. 
The total run time is $300\tau_B$. 
We choose the ratio of diffusion coefficients as $D_r/D_t\!=\!20$, although there is 
nothing special about this particular choice. 

In Fig.~\ref{fig:correlator}a we show the correlator $H(t)$ as a function of time 
for a number of different densities, the largest of which is close to the freezing 
transition for our model interaction potential. 
Aside from the strong increase of $H(0)$ with increasing density (shown in the inset), 
the most striking 
aspect of the correlator is that the decay of $H(t)$ is much faster than the timescale of 
rotational diffusion (note that time is scaled with $\tau_p$ in the figure). 
Indeed, very large values of the ratio $D_r/D_t$ would be required for the exponential 
factor in \eqref{correlator_int} to significantly influence the decay of $H(t)$. 
In the limit of large $D_r$ 
we obtain $H(t)=H(0)\exp(-2D_r t)$ and thus 
$v(\rho)/v_0 = 1 - H(0)D_t/(2D_r)$.  
We conclude that, provided the value of $D_r$ is not extremely large, 
the relevant relaxation process is the decorrelation of the 
tagged particle interaction force. 

In the inset to Fig.~\ref{fig:correlator}a we show the initial value, $H(0)$, as a function 
of the density.  
To check the expression \eqref{yvon_result} we have confirmed that using $g_{\rm eq}(r)$ 
from our equilibrium simulations to evaluate the r.h.s. indeed reproduces 
the $t\rightarrow 0$ limit of our dynamical $H(t)$ data. 
Moreover, we have also employed an approximate 
liquid-state integral equation theory (Percus-Yevick theory) \cite{HM} to calculate 
$g_{\rm eq}(r)$ and evaluate $H(0)$. Very good agreement of the predicted $H(0)$ with 
simulation data is obtained.  
%The Laplacian weight function in \eqref{yvon_result} 
%makes the integral rather insensitive to errors 

In Fig.~\ref{fig:correlator}b we replot the 
data on a semi-logarithmic scale, with the initial value scaled out. 
This representation makes clear that $H(t)$ is non-exponential and that the decay 
occurs more rapidly as the density is increased, in contrast to the structural 
relaxation of the system, which slows down with increasing density. 
The latter observation can be rationalized by considering that small positional changes can 
give rise to large changes in the force for closely packed particles residing in 
regions of strong interaction-force gradient. 
The fact that $H(t)$ is non-exponential is not surprising, given that it can be 
expressed in terms of the velocity autocorrelation function, a quantity which famously 
exhibits power law asymptotic behaviour (`long-time tails') \cite{klein,HM}. 
Klein {\it et al.} have shown analytically that for a dilute system of Brownian 
hard-spheres 
$\langle \F(t)\cdot \F(0) \rangle_{\rm eq, s}\sim t^{-\frac{5}{2}}$ for long times. 

\begin{figure}
\includegraphics[width=0.85\columnwidth]{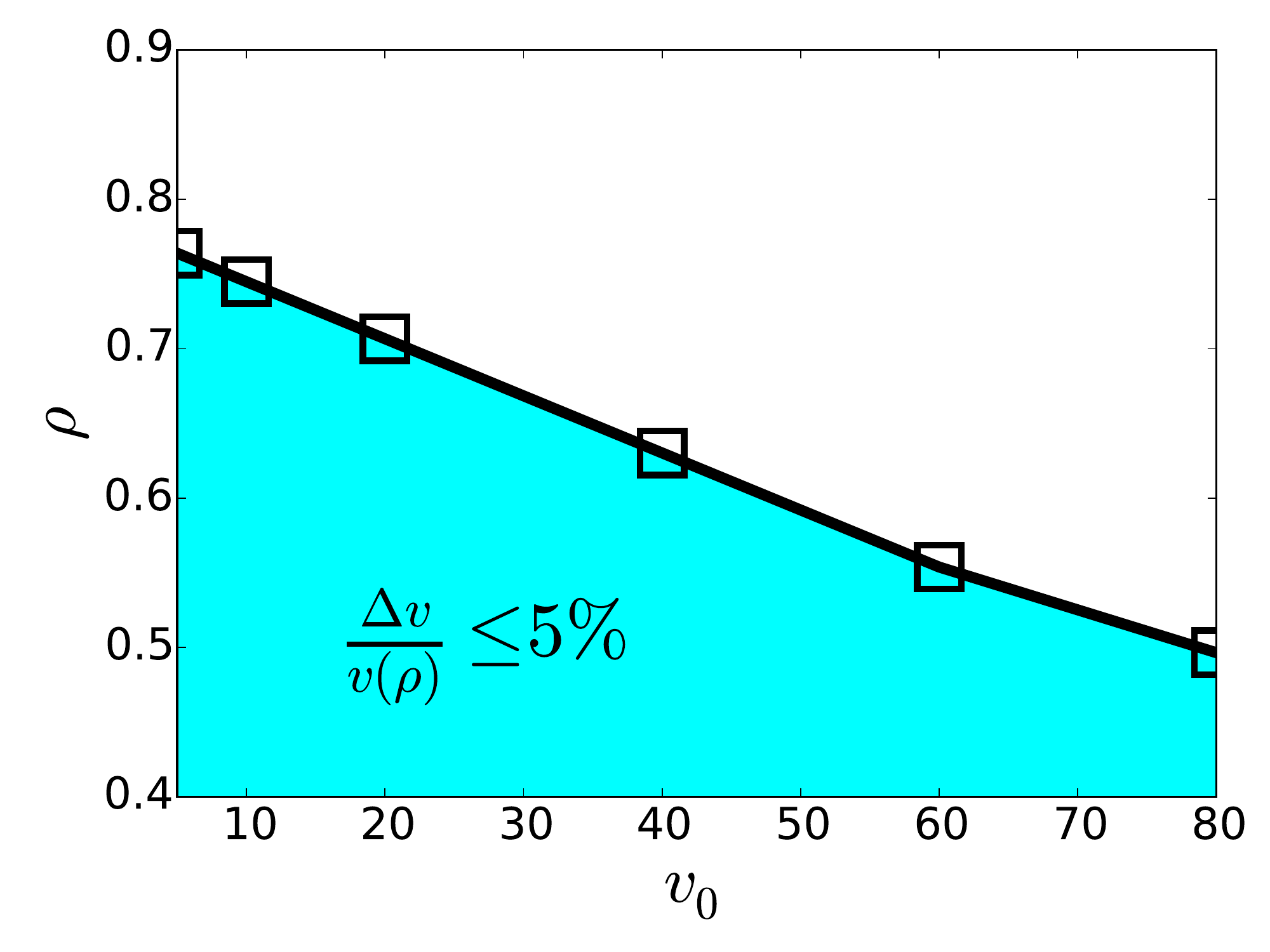}
\vspace*{0cm}
\caption{\label{fig:phase}
The region for which linear response \eqref{GK_vrho} agrees with the result of 
active-BD simulations to a relative error less than $5\%$. 
$\Delta v$ is the difference between linear response and the active BD simulation 
result. 
The breakdown of linear response is related to the onset of activity-induced 
phase separation.  
}
\label{fig1}
\end{figure}

In Fig.~\ref{fig:projected_velocity} we show simulation data for the average swim speed 
as a function of density. 
The red diamonds show the linear response prediction obtained by using the data of 
Fig.~\ref{fig:correlator} in the integral expression \eqref{GK_vrho}. This 
yields a result for $v(\rho)/v_0$ which is independent of $v_0$. 
The remaining curves show data obtained by direct evaluation of 
\eqref{swim_speed_definition} using active BD simulations at three different values 
of $v_0$.
As one might expect, deviations from linear response occur at lower density for larger values 
of $v_0$. 

The above observation can be made more concrete by estimating a region in the $(v_0,\phi)$ 
plane where linear response breaks down.
In Fig.~\ref{fig:phase} we use our simulation data to map the locus of points for which the 
error in the linear response result, relative to the full active BD simulations, equals $5\%$. 
Although the chosen criterion is somewhat arbitrary, it at least gives a visual 
impression of the range of validity of linear response within the space of our control parameters. 
The locus of points shown in Fig.~\ref{fig:phase} is correlated with the onset of strong spatial 
inhomogeneities and phase separation. 
However, an analysis of active phase separation would go beyond the scope of the present work. 
The linear response formula 
\eqref{GK_vrho} thus appears to be reliable for parameter values away from phase separation, 
but, beyond this, higher orders in $v_0$ will become important in determining $v(\rho)$. 

To summarize our main findings: we have derived a formally exact expression \eqref{nonlinear_average} 
for calculating averages in a system of interacting Brownian particles, subject to a 
time-dependent activity $v_0(t)$. From this we obtain the linear-response expression 
\eqref{linear_tindep} for a time-independent activity. 
Application of this result to calculate the average swim speed yields \eqref{GK_vrho} and 
identifies the relevant time-correlation function, $H(t)$, as given by \eqref{correlator_int}. 
We find that linear response provides an accurate account of $v(\rho)$ over a large parameter 
range, except for those regions of parameter space where phase separation occurs.  

Although we have focused our attention on the linear-response regime, our exact results could 
in principle be used to develop nonlinear theories in the spirit of 
Refs.~\cite{fuchscates2002,fuchscates2005,brader2012}, which address Brownian particles 
under external flow. 
It would also be interesting to use \eqref{nonlinear_average} to investigate the transient 
dynamics arising from time-dependent activity, but we defer this line of enquiry until an 
experimentally relevant protocol can be identified.   
Aside from using an equilibrium integral equation theory to determine $H(0)$ (inset to 
Fig.~\ref{fig:correlator}a), all of the data presented comes from BD simulation. 
A clear next step is to investigate approximations to $H(t)$ which enable predictions to 
be made from first-principles, without simulation input. 
Given the relation \eqref{H_vct} it seems likely that existing approximations to the velocity 
autocorrelation function (e.g.~projection operator approaches) could be usefully exploited.

%%%%%%%%%%%%%%%%%%%%%%%%%%%%%%%%%%%%%%%%%%%%%%%%%%%%%%%%%%%%%%%%%%%%%%%%%%%%%%%%%%%%%%%%%%%%%%%%%%%%
%%%%%%%%%%%%%%%%%%%%%%%%%%%%%%%%%%%%%%%%%%%%%%%%%%%%%%%%%%%%%%%%%%%%%%%%%%%%%%%%%%%%%%%%%%%%%%%%%%%%

\end{document}